\documentclass[usenatbib,usegraphicx]{mn2e}

\usepackage{times}
\usepackage{amssymb}

\newcommand{\aap}{A\&A}
\newcommand{\aaps}{A\&AS}
\newcommand{\aapr}{A\&AR}
\newcommand{\aj}{AJ}
\newcommand{\apj}{ApJ}
\newcommand{\apjl}{ApJ}
\newcommand{\apjs}{ApJS}
\newcommand{\mnras}{MNRAS}
\newcommand{\pasp}{PASP}
\newcommand{\ngc}{NGC\,5044}
\newcommand{\griz}{$g,\,r,\,i,\,z$\ }
\newcommand{\phn}{\phantom{1}}
\newcommand{\kms}{\mathrm{~km~s}^{-1}}
\newcommand{\msq}{\mathrm{~mag~arcsec}^{-2}}
\newcommand{\re}{\rho_\mathrm{e}}
\newcommand{\sbep}{\langle\mu_\mathrm{e}\rangle}

\defcitealias{C99}{C99}
\defcitealias{FS90}{FS90}

\begin{document}

\title[Low-luminosity galaxies in the NGC\,5044 Group]{The low-luminosity
galaxy population in the NGC\,5044 Group}

\author[Cellone \& Buzzoni]{Sergio A. Cellone$^1$
and Alberto Buzzoni$^{2}$\\
$^1$Facultad de Ciencias Astron\'omicas y Geof\'\i sicas, Universidad Nacional
de La Plata, Paseo del Bosque, B1900FWA La Plata, Argentina\\
{\sf scellone@fcaglp.unlp.edu.ar}\\
$^2$INAF - Osservatorio Astronomico di Bologna, Via Ranzani 1, 40127 Bologna,
Italy\\
{\sf buzzoni@bo.astro.it}\\
 }


\date{Accepted 2004 September 21. Received 2004 September 6; in
  original form 2004 June 23}

\maketitle

\begin{abstract}

We present multicolour imaging for a sample of 33 dwarf and
intermediate-luminosity galaxies in the field of the \ngc\ Group,
complemented with mid-resolution spectroscopy for a subsample of 13 objects.
With these data, a revised membership and morphological classification is
made for the galaxies in the sample. We were able to confirm all but one of
the ``definite members'' included in the spectroscopic subsample, which were
originally classified based on morphological criteria; however, an important
fraction of background galaxies is probably present among ``likely'' and
``possible'' members.

The presence of a nucleus could be detected in just five out of the nine
galaxies originally classified as dE,N, thus confirming the intrisic
difficulty of photographic-plate morphological classification for this kind
of objects.  Our deep surface photometry provided clear evidences for disc
structure in at least three galaxies previously catalogued as dE or dS0.
Their transition-type properties are also evident from the colour-magnitude
diagram, where they lie near the late-type galaxies locus, suggesting an
evolutionary connection between a parent disc-galaxy population and at least
part of present-day dEs.

Half a dozen new dSph candidates were also found, most of them at small
projected distances from \ngc, the central galaxy of the Group.

The \ngc\ Group appears clearly defined in redshift space, with a mean
heliocentric radial velocity, $\langle v_r \rangle = 2461 \pm 84 \kms$ ($z =
0.0082$), and a moderate dispersion, $\sigma_{v_r} = 431 \kms$.  Our
kinematical data show no luminosity segregation for early-type galaxies:
both dwarf and bright E/S0 systems show very similar velocity distributions
($\sigma_{v_r} \sim 290~\kms$), in contrast to late-type galaxies that seem
to display a broader distribution ($\sigma_{v_r} \sim 680~\kms$).

\end{abstract}



\section{Introduction \label{s_i}}

A systematic observational work to map the low surface brightness (LSB)
galaxy distribution in selected zones of the sky has been carried out by
different teams, leading to complete surveys and morphological catalogues of
some loose groups of galaxies and nearby clusters (\citealp*{BST85, KKB85,
IWO86}; \citealp{DPCDK88}; \citealp*{F89, FS90, JD97, SH97}).

These catalogues serve as a basic reference to any further analysis relying
on accurate photometry or spectroscopy of individual galaxies
\citep*[e.g.,][]{BM88, BH91, CFG94, HM94, SHP97, C99}.  However, accurate
observations of LSB galaxies still remain a quite difficult task both for
photometry and spectroscopy, especially for clusters and groups of galaxies
beyond the Virgo and Fornax Clusters.  These two nearby clusters are the
environments where, by far, most detailed studies of the dwarf and
low-luminosity galaxy population have been carried out till present,
including recent spectroscopic and photometric surveys using wide-field
detectors \citep[e.g.,][]{HKRIQ99,HIVKR99,DPJG00,KDSBH00, DBP02, SDS03}.

A complementary view can be gained through the study of smaller
groups. These have the advantage that even a moderate-sized sample would be
fairly representative of the whole group population. At the same time, depth
effects would be minimized allowing a better analysis of distance-depending
quantities.  One of these rather loose groups appears surrounding the
elliptical galaxy \ngc. The \ngc\ Group was catalogued by \citet[hereafter
FS90]{FS90}, who list 162 galaxies within $\sim 45$ arcmin of \ngc, about 80
percent of which are dwarfs ($B_{\rmn T} \ge 16$ mag).

Surface $BV$ photometry for a small sample of dwarf and intermediate
luminosity (mostly) elliptical galaxies in the \ngc\ Group (of which 6
objects in common with the present work) has been presented in
\citet[hereafter C99]{C99}, while \citet{KRP04} completed these results
with a morphological study of a brighter sample of member galaxies, 
based on $BRI$ photometry. \citet{CB01} discussed the properties of a few 
particular dwarfs, including a possible link between dEs and blue
compact dwarfs (BCD).  A systematic H\,\textsc{i} radio survey of this field 
has recently been carried out by \citet{MCK04} providing new hints
for the low-surface brightness galaxy population of this loose group. 
In this paper we further extend the \citet{CB01} analysis and
present multicolour surface photometry for 33 dwarf and 
intermediate-luminosity galaxies (plus six likely unclassified new 
members) in the field of the \ngc\ Group, along with mid-resolution 
spectroscopy for a subsample of 13 objects.

Our main goal is to clearly discriminate between members and background
objects, in order to study (in a future paper of this series) the stellar
populations and structural properties of a sample with as little background
contamination as possible. At the same time, an analysis of the kinematical
properties of the Group can be made.

We will arrange our discussion by presenting first, in Sec.~2, the
observational input. Surface photometry for each galaxy in our sample is
obtained in Sec.~3, allowing us to carry out a fully consistent
morphological (re)-classification for each object and assess --in some cases
with the combined help of spectroscopy-- its fiducial membership to the
\ngc\ Group. Nucleated and dwarf spheroidals are reviewed in some detail in
this section, and we will also present fresh data for six new galaxies,
possible Group members.

The main kinematical parameters for the Group are briefly summarized in
Sec.~4, while the photometric and structural properties of the full galaxy
sample are dealt with in Sec.~5. Here we discuss in particular the
surface-brightness vs.\ total magnitude relationship of the \ngc\ members
and pick up some notable candidates to transition-type objects. Given their
relevance in the long-standing debate about the evolutionary connection
between different dwarf-galaxy types (i.e.\ dE -- dI -- BCD), these special
objects will be further characterized in Sec.~6 comparing with the global
properties of the \ngc\ galaxy population. The main issues of our discussion
and our conclusions will finally be summarized in Sec.~7.


\begin{figure}
\includegraphics[width=\hsize]{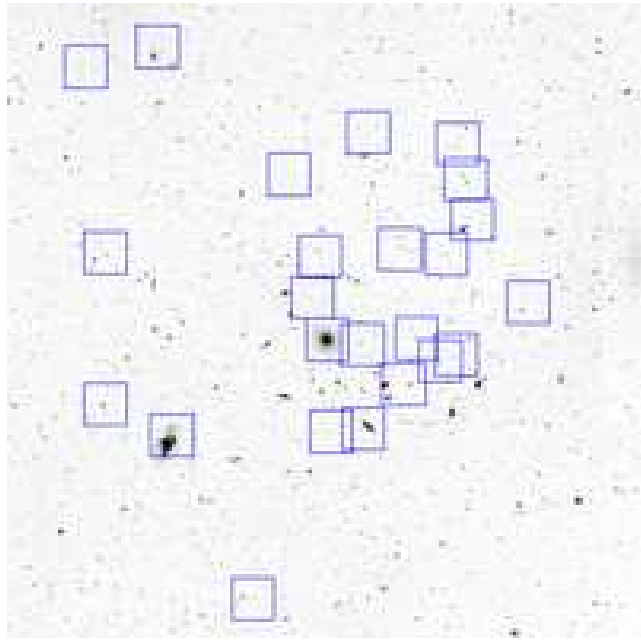}
\caption{DSS image of the \ngc\ Group, showing our observed fields. The
frame is 90 arcmin on a side, North up, East to the left.}
\label{f_dss}
\end{figure}

\section{Sample selection and observations}

A sample of dwarf and intermediate-luminosity galaxies was selected from the
\ngc\ Group Catalogue of \citetalias{FS90}, with the aim of sampling most of
the Group population between $15.0 \la B_\mathrm{T} \la 19.0$~mag (namely,
$-17 \lesssim M_B -5\,\log(h_\mathrm{o}) \lesssim
-13$~mag)\footnote{Throughout this paper, for the \ngc\ Group we will adopt
a distance modulus $(m-M) = 31.96 - 5\,\log(h_\mathrm{o})$, according to our
estimate of the mean redshift for all the known group
members (see Sec.~\ref{memb}). Note that no further correction (e.g.\ for
the Virgocentric inflow) has been applied.}  regardless of
morphology. Galaxies classified either as definite (1), likely (2), or
possible (3) members were included within our sample; only a few objects
imaged with a different telescope were excluded, since those data will be
presented elsewhere (Cellone \& Buzzoni 2004, in preparation).

 The selected fields were observed with the ESO 3.6-m telescope + EFOSC2
during two runs: April 16 -- 17, 1999, and April 29 -- May 1,
2000. Atmospheric conditions were photometric, with sub-arcsec seeing during
the first run, while in 2000 seeing was slightly poorer (FWHM $\ga 1.5$
arcsec).

\subsection{Imaging}

EFOSC2 was equipped with a Loral 2k CCD, which was $2\times 2$ binned giving
a scale of 0.32 arcsec pixel$^{-1}$. The instrument field thus covered an
square area 5.3 arcmin on a side, allowing in most cases to image more than
one galaxy within each frame. A total of 24 fields including 33
low-luminosity galaxies from \citetalias{FS90} (along with the bright SbI-II
NGC\,5054 and \ngc\ itself) were imaged in the four bands \griz of the Gunn
system \citep*{TG76, WHEH79, SGH83}.  Individual exposure times ranged from
240 sec to 720 sec; for the faintest objects two 480 sec exposures were
obtained in each band and they were summed up after all processing steps
were completed. Figure~\ref{f_dss} shows the locations of our frames on a $90
\times 90$ arcmin DSS image centred on \ngc.

Standard stars from the lists of \citet{SGH83} and \citet{J94} were also
observed during each run for calibration purposes.

Image processing was done using \textsc{iraf}\footnote{\textsc{iraf} is
distributed by the National Optical Astronomy Observatories, which are
operated by the Association of Universities for Research in Astronomy, Inc.,
under cooperative agreement with the National Science Foundation.},
complemented with a few of our own \textsc{fortran} routines.  Each frame
was bias corrected and then flat-fielded using twilight flats. The $r$ and
$i$ images showed fringe patterns which had to be corrected by subtraction
of the corresponding fringe image, scaled by an appropriate factor. Fringe
frames were constructed by median-averaging all suited science frames in $r$
and $i$, respectively; this procedure effectively removed stars and other
compact objects, although the target galaxies had to be modelled an
subtracted before the median-averaging could be done. We were able to
satisfactorily correct fringe patterns in most of the $r$ and $i$ images,
although rather large residuals ($\sim 1\%$ of the sky level) remained in a
few images.

Cosmic rays were excised using the \textsc{iraf} task \textsc{cosmicrays}.
Finally, a tilted plane was fitted to the sky background and subtracted from
each image.

\subsection{Spectroscopy}

Mid-resolution spectra were obtained for a sub-sample of 13 objects using
EFOSC2 with the grism \#8, in the wavelength range
$\lambda\lambda~4300-6400$~\AA\ at 6~\AA\ FWHM resolution.  Exposure times
ranged from 1200 sec to 4800 sec, with exposures longer than 1800 sec being
split into up to four shorter integrations. We used a long ($\sim 5$ arcmin)
slit, which allowed us, when possible, to include a second target galaxy on
the same spectrum frame of the main object. When no catalogued galaxy could
be used as second target, we tried to include any likely background galaxy
within the slit; four such objects were observed in this way. A spectrum of
the Group central galaxy, \ngc, was also obtained.

Each 2-D spectrum was bias corrected and then flat-fielded using lamp flats.
One-dimensional spectra were extracted using standard routines within
\textsc{iraf} and wavelength-calibrated by means of He-Ne-Ar lamp
spectra. Flux calibration was made with standard stars from \citet{GMCW88},
which were also observed during both runs. Finally, individual 1-D spectra of
the same object were combined.

We then used the 3--8 higher S/N absorption and/or emission lines from each of
our spectra to measure the corresponding radial velocities. Cross-correlation
with the spectrum of \ngc, used as a template, provided a confirmation, and
served as an initial guess for a few dubious, low S/N ratio spectra. All
radial velocities given along this paper are corrected to heliocentric
values.


\section{Membership and morphological re-classification} 

As usual in photographic surveys, \citetalias{FS90} used a morphological
criterion for membership classification when redshifts were not
available. This is mostly the case for dwarf ellipticals (dEs) because their
low surface brightness turns them into difficult targets for spectroscopy;
however, this LSB nature is in itself a very valuable tool that allows a
reliable membership classification, distinguishing them from background E
(i.e., high surface brightness) galaxies. Fair dEs are thus usually
classified as class 1 (definite) members, while dubious objects are assigned
to classes 2 (likely members) and 3 (possible members).  In the Virgo and
Fornax Clusters, whenever redshift information has become available for
galaxies originally assigned to class 1 on morphological bases
\citep{BST85,F89}, membership has been confirmed for an overwhelmingly large
fraction of objects \citep*[e.g.,][]{BPT93, DGHB01}, although a few striking
counterexamples are widely known \citep[e.g.: \textit{Malin 1},][]{BIMM87}.
On the other hand, spectroscopic surveys have also revealed a new population
of compact Fornax Cluster members among objects judged as background and/or
starlike on morphological bases \citep{HIVKR99, PDGJ01}.

Since the \ngc\ Group is at a distance about twice that of the Virgo
Cluster, it is necessary to test whether morphological membership
classification still holds.  Thus, our first goal was to test the
\citetalias{FS90} membership classification for our galaxies, making a more
reliable distinction between Group members and background objects. Since we
have no redshift measurement for about 60\% of our sample, we also had to
rest on morphological criteria for these objects; however, a higher spatial
resolution along with colour information allowed us to classify some dubious
cases. At the same time, we were able to give a revised morphological
classification for some objects.

\begin{figure}
\vskip 50pt
\caption{(See \emph{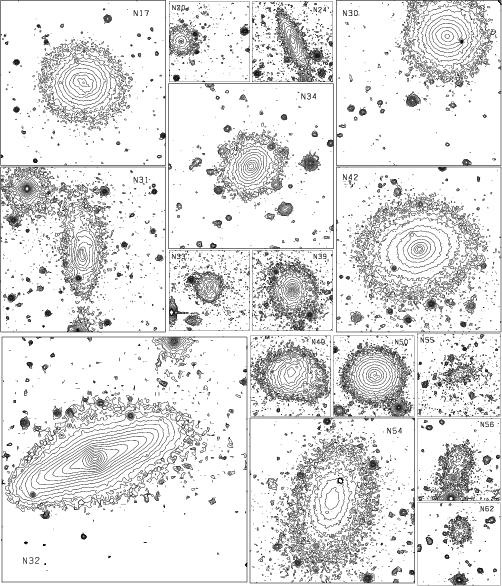} and \emph{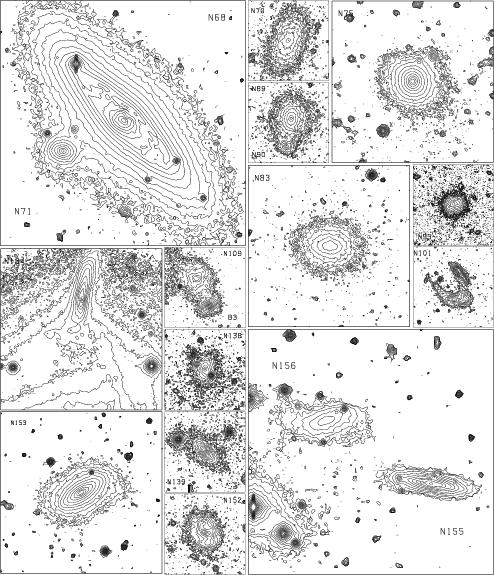}.) Contour plots from $g$
band images for all galaxies in the sample, in a standard magnitude
scale. The faintest contour corresponds to $\mu{(g)} = 26.5 \msq$, with
$\Delta \mu = 0.5 \msq$ between adjacent contours. Scale is the same for all
frames, which are either 1, 2, or 3 arcmin on a side.  North is up, East to
the left.}
\label{f_conto}
\end{figure}

Figure~\ref{f_conto} shows $g$ band contour plots for each galaxy in our
sample. (``True'' colour images from combined $g$, $r$, and $i$ frames are
available only in the electronic version.)  For this graphical presentation,
all images were sky-subtracted, median filtered and converted to the standard
magnitude scale. The faintest contour corresponds to $\mu(g) = 26.5 \msq$,
with $\Delta \mu = 0.5 \msq$ between adjacent contours. Scale is the same for
all frames, which range from 1 to 3 arcmin on a side. Note the wide variety
of sizes and morphologies; these will be discussed later.

From our spectroscopic data, we obtain $v_r=2710 \pm 40 \kms$ for \ngc, in
coincidence with published values \citep[$v_r=2704 \kms$,][]{HDLT83, ssrs}.
For the dwarf elliptical galaxies N42 and N50\footnote{In what follows, the
prefix ``N'' stands for catalogue number in \citetalias{FS90}}, we confirm,
within the errors, the preliminary $v_r$ values given by \citetalias{C99}.

\begin{figure}
\includegraphics[width=\hsize]{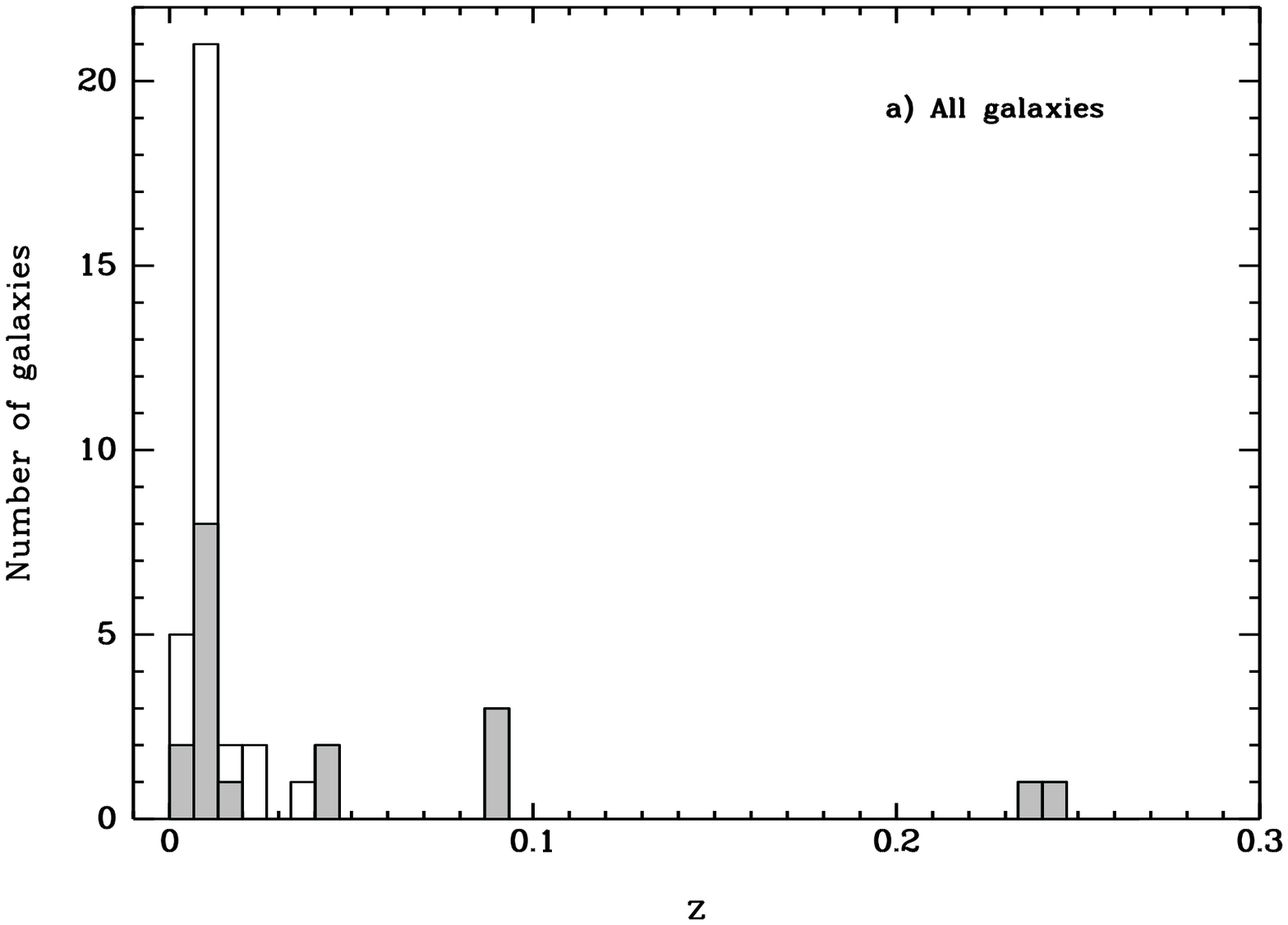}
\includegraphics[width=\hsize]{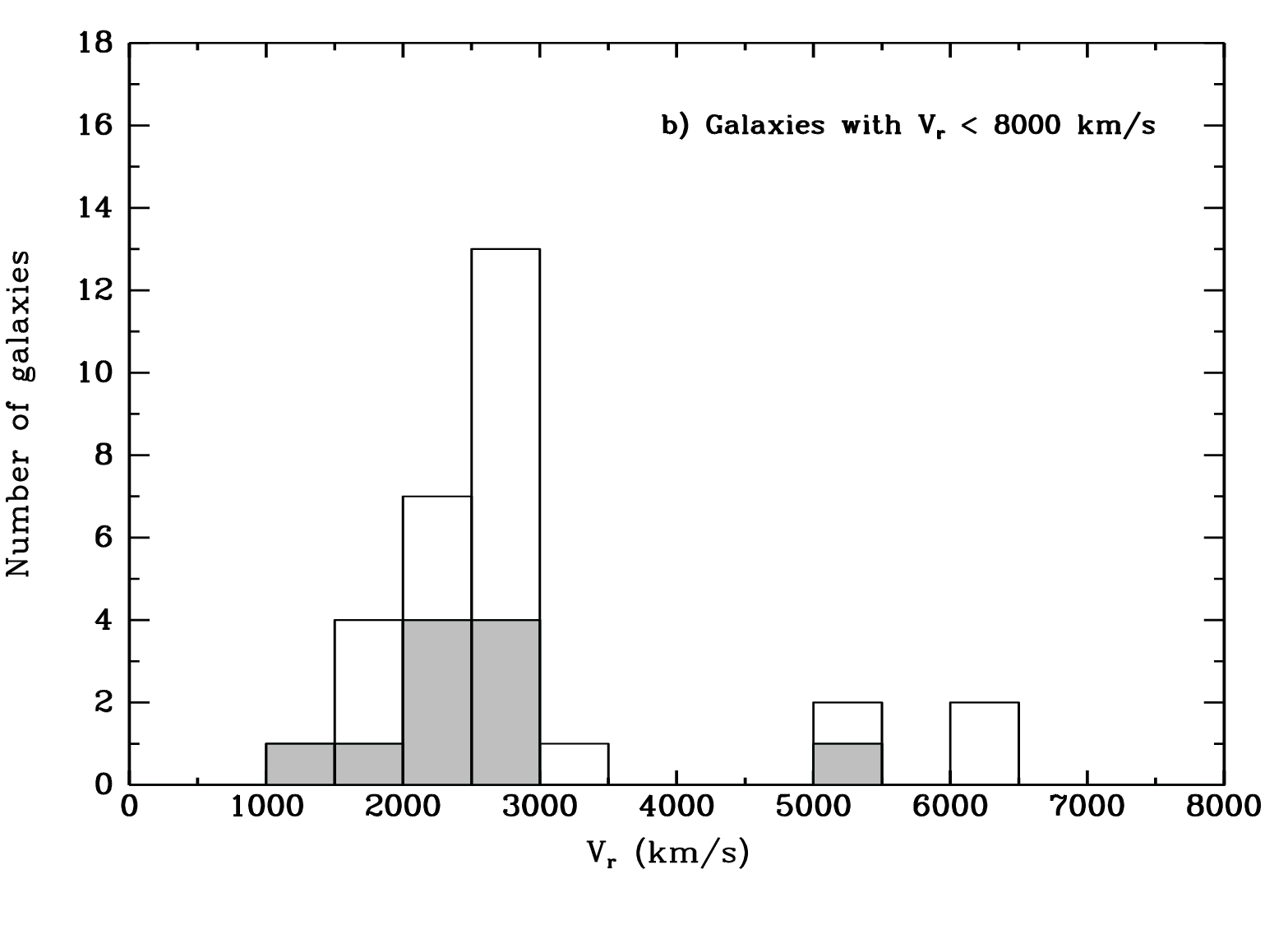}
\caption{Radial velocity distribution for galaxies within $\sim 45$ arcmin
from \ngc. Filled bars correspond to new data presented in this paper, while
empty ones show data from the literature. (a) All galaxies; (b) galaxies
with $v_r < 8000 \kms$.}
\label{f_hvr}
\end{figure}

Figure~\ref{f_hvr} shows the radial velocity distribution for galaxies with 
spectroscopic data within $\sim 45$ arcmin from \ngc; filled bars correspond
to new data presented in this paper,%
\footnote{Including the dwarf N29, from  Buzzoni \& Cellone (2004, in
preparation)} 
while the empty ones show data from the literature, provided
by NED.  Panel \emph{a}) spans the whole velocity range, while \emph{b}) 
singles out those objects with $v_r < 8000 \kms$.

The \ngc\ Group appears clearly defined in redshift space, with a mean
heliocentric radial velocity $\langle v_r \rangle = 2461 \pm 84 \kms$ 
($z = 0.0082$) and a
moderate dispersion $\sigma_{v_r} = 431 \kms$. A gap is evident between
$\sim 3500$ and $\sim 5000 \kms$. Inclusion of the four galaxies with $5400
\la v_r \la 6400 \kms$ as members would raise the Group velocity dispersion
to an implausibly large 1262 km s$^{-1}$. Three of these objects appear
concentrated in the south-east quadrant of the Group, so they might belong
to a background structure.  In what follows, we shall thus
consider all galaxies with $v_r > 3800 \kms$ (i.e., $v_r \ga \langle v_r
\rangle + 3 \sigma_{v_r}$) to be background objects.

\begin{table}
\caption{Membership and morphological re-classification}
\label{t_clas}
\begin{tabular}{@{}lllc@{}}
\hline
FS90 Nr.   & Morph. & $v_r$~~~~   & Member\\
           &        & $\kms$  &  \\
\hline
\noalign{\smallskip}
\multicolumn{4}{c}{Definite members ($m=1$)} \\
~~\phn 20  & dE,N            &           & \textsc{yes}\\
~~\phn 30  & \textsl{dE}     & \phn2411  & \textsc{yes}\\
~~\phn 32  & S0              & \phn2795$^a$ & \textsc{yes}\\
~~\phn 34  & \textsl{dE}     & \phn2661  & \textsc{yes}\\
~~\phn 42  & \textsl{dE/dS0} & \phn2462  & \textsc{yes}\\
~~\phn 49  & ImIII           & \phn1499  & \textsc{yes}\\
~~\phn 50  & dE,N pec/BCD    & \phn2392  & \textsc{yes}\\
~~\phn 54  & dE(Huge)        &     	     & \textsc{yes}\\
~~\phn 56  & \textsl{dSph}   &     	     & \textsc{yes}\\
~~\phn 62  & \textsl{dSph}   &     	     & \textsc{yes}\\
~~\phn 68  & Sab(s)          & \phn1887$^b$ & \textsc{yes}\\
~~\phn 70  & \textsl{dE,N/dI}&     	     & \textsc{yes}\\
~~\phn 71  & \textsl{d:E,N?} &     	     & \textsc{yes}\\
~~\phn 75  & \textsl{dE}     & \phn1831  & \textsc{yes}\\
~~\phn 83  & dE              &     	     & \textsc{yes}\\
~~\phn 84$^c$ & E            & \phn2710  & \textsc{yes}\\
~~\phn 89  & dE,N            &     	     & \textsc{yes}\\
    ~~109  & \textsl{Sdm}    & \phn5409  & \textsc{no}\\
    ~~134  & Sm(interacting) &     	     & \textsc{yes}\\
 ~~137$^d$ & Sb(s)I-II(int.) & \phn1743$^b$ & \textsc{yes}\\
    ~~153  & d:S0            & \phn2816  & \textsc{yes}\\
    ~~155  & \textsl{Sab}    & \phn2922  & \textsc{yes}\\
    ~~156  & \textsl{dE/dI}  &           & \textsc{yes}\\

\noalign{\smallskip}
\multicolumn{4}{c}{Likely members ($m=2$)} \\
~~\phn 17  & \textsl{S0pec } & \phn2682  & \textsc{yes}\\
~~\phn 24  & \textsl{Sm}     &           & \textsc{no?}\\
~~\phn 55  & \textsl{dE/dI}  &           & \textsc{yes}\\
~~\phn 90  & \textsl{Sc}     &           & \textsc{no}\\
    ~~138  & dE,N            &           & \textsc{yes}\\
    ~~152  & \textsl{Sd}     &     13580 & \textsc{no}\\
\noalign{\smallskip}
\multicolumn{4}{c}{Possible members ($m=3$)} \\
~~\phn 31  & \textsl{Im}     &           & \textsc{yes}\\
~~\phn 33  & \textsl{SB?c}   &     13480 & \textsc{no}\\
~~\phn 39  & \textsl{E or cD}&     27390 & \textsc{no}\\
~~\phn 93  & bSp or Sm?      &           & \textsc{yes?}\\
    ~~101  & interacting     &           & \textsc{no}\\
    ~~139  & \textsl{E}      &           & \textsc{no?}\\
\hline
\noalign{\smallskip}
\noalign{$a$: $v_r$ from \citet{FWC-92}}
\noalign{\smallskip}
\noalign{$b$: $v_r$ from \citet{HDLT83}}
\noalign{\smallskip}
\noalign{$c$: NGC\,5044}
\noalign{\smallskip}
\noalign{$d$: NGC\,5054}
\end{tabular}
\end{table}


Table~\ref{t_clas} shows the results of our re-classification. Column~1 is
the galaxy number in \citetalias{FS90}, and column~2 gives our morphological
re-classification. Whenever this last differed from the original one, a
slanted typography was used. Column 3 gives the heliocentric radial velocity
(when available); finally, column~4 gives our membership classification
(question marks were used for a few dubious cases). Radial velocity
uncertainties for our data in Table~\ref{t_clas} typically spanned the range
$50 \kms\lesssim \Delta\,v_r \lesssim 100 \kms$, except for N109
($\Delta\,v_r \simeq 250 \kms$).

Among ``definite members'' ($m=1$) with new redshift data (10 objects), only
N109 ($v_r=5409 \kms$) lies in the near background. Classified as
ImV or dE,N?, this object is in fact a late-type spiral galaxy (see
Fig.~\ref{f_conto}). The remaining 9 galaxies are LSB dwarfs; their
confirmation as Group members gives further support to the accuracy of
morphological membership assignment for this kind of objects.

Aside from N109, no other ``definite'' member was re-classified as
background on morphological bases. Most of these objects are dEs which can
be safely considered as members. The Sm galaxy N134, in turn, is interacting
with NGC\,5054, and hence it must belong to the Group.


%


%

Regarding ``likely'' ($m=2$) and ``possible'' members ($m=3$), we present
radial velocities for two objects within each class. Only N17 ($m=2$) was
confirmed as a Group member, although with a different morphological
classification. The other three objects turned out to be in the background:
N33 and N152 are late-type spirals, while N39 is a luminous elliptical ($M_V
\simeq -22.3$). Judging from the ``cuspy'' shape of its surface brightness
profile (see Sec.~\ref{s_SBP}) N39 is probably the brightest galaxy in a
background cluster \citep[e.g.,][]{GLCP96}, to which the object B1 (see
Table~\ref{t_back}) also belongs (note that object B3 is at the same
redshift, however it is at $\sim 1^\circ$ projected distance from N39). Its
failure as a Group member is a new confirmation of the extreme rarity of
M32-type compact ellipticals \citep{ZB98, DGHB01}.

Based on morphological criteria, we also reject N90, which we classify as Sc
instead of dE,N (see Fig.~\ref{f_conto}; faint but well defined blue spiral
arms with a few H\,\textsc{ii} regions are discernible on the colour
picture). The interacting pair N101 is also probably in the background,
while no definite assertion can be made about the late-type spiral N24, the
high surface brightness compact object N93, and the E or M32-type
N139. We only retain N31 and N55 as Group members; the first one because of
its resolution into star forming regions, and the second one because of its
dwarf spheroidal (dSph) morphology.

As expected, all four objects included in our spectroscopy as ``bonus'' 
targets on the slit were confirmed as background galaxies 
(see Table~\ref{t_back} and Fig.~\ref{f_hvr}; typical redshift
uncertainties are $\Delta z \lesssim 0.001$).

\subsection{Nucleated dwarfs}

We observed 9 galaxies classified as dE,N (i.e., nucleated dwarf elliptical)
in \citetalias{FS90}. Despite the good seeing conditions during our
observing runs, we could not detect any nucleus in N30 and N34
\citepalias[see also][]{C99}; in neither case there is any luminosity
enhancement in the inner regions of their surface brightness profiles (see
sect.~\ref{s_SBP}).  In turn, N42 and N75 show a ``cuspy'' profile or
bulge-type inner component, but no distinct compact central component to be
regarded as a nucleus.

Hence, four out of the nine dwarfs originally classified as nucleated within
our (small) sample have in fact no nucleus. Qualitatively similar results
for small samples of Fornax Cluster dwarfs have been obtained previously
\citep{CFG94, DBP02}. In fact, \citet{CFG94} noted that the two largest
photographic surveys in the Fornax Cluster (namely, \citealp{DPCDK88} and
\citealp{F89}) disagree with each other in over 50\% of the cases in
classifying dwarf ellipticals as nucleated or non-nucleated. A high degree
of coincidence is achieved, instead, between different CCD studies
\citep[e.g.,][]{C83, CFG94}.

We judge this problem is a limitation inherent to photographic
classification, and should be taken in mind when analysing the properties of
nucleated vs.\ non-nucleated dwarfs, since differences between both
subclasses have been reported regarding their structure, spatial
distributions, and stellar populations \citep[e.g.,][]{FS89,RT94,RS03}. A
clearer identification of nucleated dwarfs should also help to better
understand their possible connection with intra-cluster globular clusters
and ultra-compact dwarfs \citep*{BCFD03, MHI04}.

\subsection{Candidate dSph galaxies
\label{s_dSph}}

\begin{figure}
%
%
\vskip 50pt
\caption{(See \emph{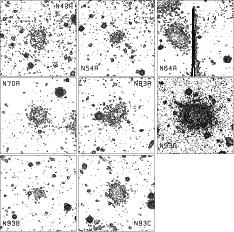}.) The six new dSph candidates (plus the imaged
field for galaxies N49A and N83A, already reported in
\citetalias{C99}). Each image is 1 arcmin on a side, with North up and East
to the left. Contour representation is the same as for Fig.~\ref{f_conto}.}
\label{f_dSph}
\end{figure}

Six new very LSB galaxies ($\langle\mu_\mathrm{e}\rangle{\rm (g)} \sim 25.5
\msq$) were detected by visual inspection on our frames. Their morphologies
suggest that they are previously non-catalogued \ngc\ Group members; we
named them by appending a capital letter to the name of the nearest Group
member. Table~\ref{t_dSph} lists their equatorial coordinates along with our
classification. Contour plots for these new galaxies are shown in
Fig.~\ref{f_dSph}, along with two similar objects (namely, N49A and N83A)
first reported in \citetalias{C99} and re-imaged here.

Although we did not set any {\it a priori} detection criteria, a subsequent
analysis shows that we were able to detect objects with isophotal $g$ band
radii (at $\mu(g) = 27 \msq$) larger than $\sim 5$ arcsec. All eight
galaxies are very LSB, with central surface brightnesses fainter than $\sim
24.5 \msq$. N93A is a somewhat extended object (isophotal radius $\simeq
14.5$ arcsec) with a lumpy appearance and a very shallow surface brightness
profile, while N93C looks more like a faint nucleated dE.

\begin{table}
\caption{Coordinates and redshifts of background ``bonus'' galaxies.}
\label{t_back}
\begin{tabular}{@{}lccc@{}}
\hline
Name & $\alpha_\mathrm{J2000}$ & $\delta_\mathrm{J2000}$ & $z$ \\
\hline
B1 & $13^{\rm h}\, 14^{\rm m}\, 19.7^{\rm s}$ & $-16\degr\, 10'\, 30''$ &
0.097 \\
B2 & $13^{\rm h}\, 15^{\rm m}\, 01.9^{\rm s}$ & $-16\degr\, 22'\, 16''$ &
0.282 \\
B3 & $13^{\rm h}\, 16^{\rm m}\, 07.2^{\rm s}$ & $-17\degr\, 00'\, 14''$ &
0.096 \\
B4 & $13^{\rm h}\, 17^{\rm m}\, 42.0^{\rm s}$ & $-16\degr\, 10'\, 05''$ &
0.277 \\
\hline
\end{tabular}
\end{table}
%
%
\begin{table}
\caption{Coordinates and classification of new very LSB galaxies.}
\label{t_dSph}
\begin{tabular}{@{}lccl@{}}
\hline
Name & $\alpha_\mathrm{J2000}$ & $\delta_\mathrm{J2000}$ & Morph.\\
\hline
N54A & $13^{\rm h}\, 14^{\rm m}\, 42.8^{\rm s}$ & $-16\degr\, 11'\, 18''$ &
dSph \\
N64A & $13^{\rm h}\, 14^{\rm m}\, 48.9^{\rm s}$ & $-16\degr\, 30'\, 30''$ &
dSph \\
N70A & $13^{\rm h}\, 14^{\rm m}\, 56.7^{\rm s}$ & $-15\degr\, 52'\, 31''$ &
dSph \\
N93A & $13^{\rm h}\, 15^{\rm m}\, 36.9^{\rm s}$ & $-16\degr\, 19'\, 22''$ &
dSph/dI \\
N93B & $13^{\rm h}\, 15^{\rm m}\, 30.9^{\rm s}$ & $-16\degr\, 19'\, 22''$ &
dSph \\
N93C & $13^{\rm h}\, 15^{\rm m}\, 35.9^{\rm s}$ & $-16\degr\, 18'\, 43''$ &
dE,N \\
\hline
\end{tabular}
\end{table}
%

Except for N70A ($d=31.4$ arcmin), the new dSphs, along with the faintest
catalogued dEs (N55, N56, and N62), lie within a projected distance $d=18$
arcmin ($130\, h_\mathrm{o}^{-1}$ kpc) from \ngc. Furthermore, N55 and N64A
are very close (a few arcmin) to the bright SBa NGC\,5035; in fact, N64A was
not evident until the halo of NGC\,5035, as well as the wings of a saturated
foreground star, were subtracted. Although the number of objects is low for
statistically significant conclusions, their apparently clustered
distribution would be similar to that of Local Group dSphs, which tend to be
found near the Milky Way and M31. This contrasts with \citet{KDSBH00}, who
found a very LSB galaxy distribution in the Fornax Cluster less concentrated
than the bright galaxies (but, again, see \citealp*{HMI03}, for an
opposite conclusion on the same cluster, in line with our results).


\section{Group kinematics} \label{memb}

Table~\ref{t_vr} shows mean heliocentric radial velocities and dispersions
for the galaxies in three different morphological classes: bright early-type
galaxies (E--S0), early type dwarfs (dE--dS0), and late-type objects
(Sa--Im), as well as for all Group members. For ellipticals an S0s, the
boundary between giants and dwarfs was set at $B_\mathrm{T} = 15.5$ mag
($M_B = -16.5 - 5\,\log(h_\mathrm{o})$), while the Sa--Im set encompasses
both bright and dwarf objects.

Late-type galaxies seem to have a sensibly broader distribution than both
dwarf and bright early-type galaxies.  The intrinsically small number of
members of the \ngc\ Group prevents, in our case, to achieve any firm
quantitative conclusion in this regard; however, an $F$ test on the velocity
dispersion data of Table~\ref{t_vr} confirms that late- and early-type
galaxies are differently concentrated at a moderately high 93\% confidence
level.

Focusing on the early-type galaxies, and judging from their respective means
and dispersions, it is evident that the velocity distributions of dwarf and
bright objects are indistinguishable. At least for the magnitude ranges
spanned by both sets, there is no hint for the luminosity segregation
that shows up in the kinematic data of early-type galaxies in Virgo, which
has been attributed to the merging of two sub-clumps within the Virgo core
\citep{BPT93}. It has also been shown that dwarf irregulars
have a broader velocity distribution than bright late-type galaxies in two
nearby groups \citep{CFCQ97}; however, our late-type subsample is too small for
further subdivision, so we cannot test this result. 

Weighting the data by luminosity does not change things much, except for the
fact that \ngc\ and NGC\,5054, both being $\sim 1.5$ mag brighter than the
next brightest galaxy, tend to bias the weighted means toward higher and
lower velocities, respectively.  Note that the velocity of \ngc\ is nearly
$1 \sigma$ larger than the Group and bright-E means; this agrees with
\citet{DJFD94} who find evidences from X-ray data that \ngc\ has a residual
velocity with respect to the centre of the Group potential.

Our results are thus in qualitative agreement with the well known
kinematical morphology segregation in clusters \citep[e.g.,][and references
therein]{BKTA02}. Statistical studies have also found kinematical
morphology segregation in groups; although there is some discrepancy
regarding its significance, this seems to be mostly so for poor groups
with velocity dispersions substantially lower than the \ngc\ Group
\citep*{GRMM03,LGLS04}.  On the other hand, luminosity segregation is found
for galaxies $\sim 3.5$ mag brighter than our $B_\mathrm{T} = 15.5$ mag
giant -- dwarf boundary. The lack of luminosity segregation in our
early-galaxies data appears to support the \citet{LGLS04} finding that
morphology (or colour index) is the primary parameter in defining dynamical
properties of galaxies in groups, while luminosity segregation is mostly due
to the luminosity -- morphology correlation.


\section{Photometric and structural properties}

\subsection{Surface brightness profiles} \label{s_SBP}

For each galaxy we obtained the surface brightness profile (SBP) using the
\textsc{iraf} task \textsc{ellipse}. Compact foreground or background objects
which could disturb the isophote-fitting algorithm were previously masked
out. When necessary, the haloes of large neighbour galaxies (e.g.,
NGC\,5035) were modelled and subtracted; the same was done with the outer
wings of the PSFs of bright, saturated stars. Several non-saturated bright
stars were also subtracted using \textsc{daophot}.

SBPs were obtained from the $g$ band images, using a three-step procedure.
An inner region (semi-major axis $a < a_1$) was defined as that where
the error bars of the centre coordinates remained below 0.25 pixels (0.08
arcsec); within this region, all the ellipse parameters were allowed to vary
freely.  A second limiting semi-major axis ($a_2$) was set at the point
where the isophote intensity dropped below a value twice its own rms; this
point typically corresponds to a surface brightness $\mu{(g)} \approx 26
\msq$. For this middle region ($a_1 < a < a_2$) the centre was fixed and the
fitting algorithm continued with variable ellipticity and position
angle. From $a_2$ outwards, all ellipse parameters remained fixed. This
procedure prevented against ``isophote wandering'' and erratic changes in
ellipticity and position angle at low surface brightness levels. However,
for a few faint and/or compact galaxies, the middle region practically
vanished, while for the faintest dwarfs (see Sect.~\ref{s_dSph}) no ellipse
fitting was possible at all, so we had to force fixed elliptical or circular
apertures.  For each object, the $g$ band elliptical parameters were used to
obtain the profiles in the remaining bands ($r$, $i$, $z$).

Final steps involved: a fine-tuning of the sky level (typically a few adu)
by checking the flatness of the growth curve far from the galaxy centre,
conversion of semi-major axis ($a$) to equivalent radius\footnote{We use
$\rho$ for the equivalent radius to avoid confusion with the $r$ passband.}
$\rho=\sqrt{ab}$ in arcsec, and transformation of instrumental intensities
to surface brightness ($\mu$) in standard magnitudes per square arcsec.

\begin{table}
\caption{Kinematic properties.}
\label{t_vr}
\begin{tabular}{@{}lrcccc@{}}
\hline
     &     & \multicolumn{2}{c}{no weights} & \multicolumn{2}{c}{$L$
     weighted}  \\
Type & Nr. & $\langle v_r \rangle$ & $\sigma_{v_r}$ & $\langle v_r \rangle$
& $\sigma_{v_r}$\\
     &      &  \multicolumn{2}{c}{km s$^{-1}$} & \multicolumn{2}{c}{km
     s$^{-1}$}  \\
\hline
E--S0   & 9 & $2488 \pm \phn 96$  & 287 & $2590 \pm \phn 87$ & 262 \\
dE--dS0 & 9 & $2487 \pm \phn 98$  & 294 & $2493 \pm \phn 97$ & 292 \\
Sa--Im   & 8 & $2404 \pm 241$ & 681 & $1982 \pm 245$ & 693 \\
All    & 26 & $2461 \pm \phn 84$  & 431 & $2281 \pm \phn 98$ & 501 \\
\hline
\end{tabular}
\end{table}

\subsubsection{Model independent parameters}

Most of our $g$ band SBPs can be traced out to a surface brightness $\mu{(g)}
\simeq 28 \msq$, although photometric errors get too large for intensities
below one half the isophote rms, or about $\mu{(g)} \ga 27 \msq$. We thus
chose the $\mu{(g)} = 27.0 \msq$ isophote to measure isophotal radii
($\rho_{27}$), magnitudes ($g_{27}$), and mean surface brightnesses
($\mu_{27}(g)$). The $g$ band $\rho_{27}$ was then used to measure
magnitudes and surface brightnesses in the remaining bands ($r$, $i$, $z$),
so that all these parameters refer to the same physical radius.

The model independent effective radius ($\re$) was obtained from each growth
curve as the radius encompassing half the luminosity within the $\rho_{27}$
isophote. It is clear that this is an underestimation of the actual
half-light radius; however, we shall show that our isophotal magnitudes are
fairly good approximations to the total brightnesses that one would obtain
by integrating the observed profile to infinite radius. Consequently, the
model independent effective surface brightness ($\mu_\mathrm{e}$) was
obtained as the surface brightness at the $\rho = \re$ isophote, and the
mean effective surface brightness as
$\langle\mu_\mathrm{e}\rangle = g_{27} + 5 \log(\re) + 1.995$.

\subsubsection{SBP fitting}

We fitted the $g$ band SBPs with a \citet{S68} law: $\mu (\rho) = \mu_0 +
1.086 \left({\rho / \rho_\mathrm{o}}\right)^n$, where $\mu_0$ is the central
surface brightness, $\rho_\mathrm{o}$ is the pseudo scale-length, and $n$ is
the S\'ersic index which governs the SBP shape. Note that we use $n$ as the
exponent in the S\'ersic law, as usual for dwarf galaxies, instead of $1/n$.

It is known that fitting performances are strongly sensitive to the
appropriate match of the SBP.  Seeing plays an important role for small
$\rho$, while noise and sky uncertainties affect on the contrary at large
distances \citep{D97, C99, KIvDF00}.  All fits were therefore carried out
over a region comprised between an inner radius about twice the seeing FWHM
(namely, $\rho \sim 2$~arcsec), and an outer radius assuring a sky-corrected
intensity at least equal to its rms (this corresponds roughly to $\mu{(g)}
\simeq 26.5 \msq$).

The flexibility of S\'ersic's law allowed us to obtain satisfactory fits for
most of our dwarfs, despite their very different shapes. This is true even
for the few dwarfs in our sample with evidences for current star formation
(N24, N31, N49).

A new set of photometric parameters, but now model dependent ones
($\rho_{27}^\mathrm{S}$, $g_{27}^\mathrm{S}$,
$\langle\mu_\mathrm{e}^\mathrm{S}\rangle$, etc.), were then calculated from
the S\'ersic fits; when necessary, we identify them with a superscript ``S''
to distinguish them from their model-independent counterparts. Total
magnitudes were eventually estimated by integrating the S\'ersic profile:
$g_\mathrm{T} = \mu_\mathrm{o} - 2.5 \log\left(2\pi\rho_\mathrm{o}^2\right)
- 2.5 \log\left[{\Gamma\left({2/n}\right) /n}\right]$. Table~\ref{t_photpar}
lists the relevant photometric parameters for galaxies in our sample.

\begin{table*}
\caption{Observed potometric parameters}
\label{t_photpar}
\begin{tabular}{@{}lrrrrrrrrr@{}}
\hline
Name & $\rho_{27}$ & $\rho_\mathrm{e}$ & $g_{27}$ & $\mu_{27(g)}$ &
$\langle\mu_\mathrm{e}\rangle$ & $\mu_{0}$ & $\rho_0$ & $n$ & $g_\mathrm{T}$
\\
     & \multicolumn{2}{c}{arcsec} & mag & \multicolumn{3}{c}{$\msq$} & arcsec
& & mag \\
\hline
\phn 17 & 33.1 &  9.6 & 15.55 & 24.40 & 22.47 & 21.11 &  4.35 &  0.84 & 15.52 \\
\phn 20 & 12.4 &  4.9 & 18.70 & 25.42 & 24.17 & 23.59 &  4.68 &  1.12 & 18.44 \\
\phn 24 & 16.7 &  5.0 & 16.91 & 24.26 & 22.42 & 21.58 &  3.92 &  1.19 & 16.91 \\
\phn 30 & 35.4 & 10.9 & 15.51 & 24.50 & 22.68 & 20.87 &  4.17 &  0.80 & 15.24 \\
\phn 31 & 29.3 &  8.4 & 16.33 & 24.91 & 22.95 & 20.93 &  1.71 &  0.61 & 16.22 \\
\phn 32 & 58.0 &  7.5 & 12.83 & 22.89 & 19.19 & 16.95 &  0.97 &  0.55 & 12.93 \\
\phn 33 & 13.4 &  4.0 & 17.57 & 24.44 & 22.59 & 21.92 &  3.47 &  1.21 & 17.54 \\
\phn 34 & 25.9 &  6.1 & 16.28 & 24.59 & 22.20 & 18.87 &  0.28 &  0.44 & 16.04 \\
\phn 39 & 21.8 &  3.8 & 16.49 & 24.42 & 21.38 & 13.66 &  0.00028 &  0.22 & 16.23 \\
\phn 42 & 44.0 & 13.7 & 15.25 & 24.72 & 22.93 & 22.97 & 14.60 &  1.18 & 15.43 \\
\phn 49 & 25.3 &  8.2 & 15.75 & 24.01 & 22.33 & 21.58 &  6.96 &  1.30 & 15.78 \\
\phn 49A & 5.9 &  3.1 & 20.94 & 26.02 & 25.37 & \dots & \ldots & \dots & \dots \\
\phn 50 & 26.0 &  7.1 & 15.21 & 23.52 & 21.47 & 20.40 &  4.52 &  1.04 & 15.19 \\
\phn 54 & 42.6 & 18.1 & 15.95 & 25.34 & 24.24 & 22.95 & 11.10 &  1.01 & 15.75 \\
\phn 54A & 5.5 &  3.2 & 21.31 & 26.27 & 25.86 & 25.64 &  4.82 &  1.76 & 20.92 \\
\phn 55 &  9.4 &  4.4 & 19.64 & 25.74 & 24.83 & 22.76 &  0.98 &  0.61 & 19.18 \\
\phn 56 & 13.1 &  6.0 & 18.94 & 25.77 & 24.82 & 24.35 &  6.88 &  1.48 & 18.72 \\
\phn 62 &  9.5 &  5.0 & 19.80 & 25.93 & 25.29 & 24.79 &  6.43 &  1.26 & 19.13 \\
\phn 64A & 9.9 &  4.8 & 19.48 & 25.69 & 24.87 & 24.62 &  6.10 &  1.92 & 19.43 \\
\phn 68 & 82.3 & 16.3 & 12.29 & 23.11 & 20.34 & 17.64 &  2.11 &  0.58 & 12.23 \\
\phn 70 & 22.5 &  9.6 & 17.25 & 25.26 & 24.17 & 23.37 &  7.92 &  1.15 & 17.12 \\
\phn 70A & 7.7 &  4.3 & 20.50 & 26.17 & 25.67 & 25.43 &  6.23 &  1.83 & 20.17 \\
\phn 71 & 28.0 &  6.8 & 16.43 & 24.90 & 22.60 & 18.78 &  0.17 &  0.40 & 16.25 \\
\phn 75 & 29.4 &  6.7 & 15.40 & 23.98 & 21.53 & 22.21 &  9.85 &  1.42 & 15.76 \\
\phn 83 & 28.9 & 11.0 & 16.70 & 25.25 & 23.90 & 22.75 &  6.38 &  0.92 & 16.55 \\
\phn 83A &  9.7 &  5.2 & 20.07 & 26.24 & 25.63 & 24.86 &  4.81 &  1.07 & 19.59 \\
\phn 89 & 20.5 &  6.4 & 17.47 & 25.28 & 23.49 & 21.92 &  2.32 &  0.71 & 17.15 \\
\phn 90 &  8.2 &  3.2 & 18.97 & 24.79 & 23.49 & 22.67 &  2.27 &  1.00 & 18.88 \\
\phn 93 & 12.2 &  3.9 & 17.58 & 24.25 & 22.53 & 21.53 &  2.68 &  1.09 & 17.55 \\
\phn 93A & 14.5 &  8.0 & 18.91 & 25.97 & 25.42 & 25.14 & 10.60 &  2.16 & 18.81 \\
\phn 93B &  4.6 &  2.5 & 21.59 & 26.15 & 25.54 & 25.01 &  2.93 &  1.84 & 21.39 \\
\phn 93C &  8.6 &  4.0 & 19.81 & 25.73 & 24.81 & 24.40 &  4.36 &  1.50 & 19.77 \\
 109 & 16.1 &  6.5 & 17.66 & 24.95 & 23.72 & 22.57 &  4.09 &  0.96 & 17.44 \\
 134 & 36.6 &  5.9 & 15.23 & 24.28 & 21.09 & 19.22 &  1.43 &  0.64 & 15.12 \\
 138 & 15.4 &  5.2 & 18.15 & 25.33 & 23.72 & 22.39 &  2.49 &  0.82 & 17.93 \\
 139 & 15.5 &  3.2 & 17.22 & 24.41 & 21.75 & 19.13 &  0.36 &  0.54 & 17.16 \\
 152 & 18.2 &  6.1 & 16.72 & 24.26 & 22.64 & 22.14 &  6.46 &  1.62 & 16.72 \\
 153 & 46.9 & 11.1 & 14.80 & 24.40 & 22.02 & 19.40 &  1.20 &  0.53 & 14.71 \\
 155 & 23.4 &  6.2 & 16.19 & 24.28 & 22.17 & 20.52 &  2.28 &  0.79 & 16.13 \\
 156 & 31.8 & 13.5 & 16.97 & 25.73 & 24.61 & 23.52 &  9.15 &  1.05 & 16.81 \\
 ~B3 & 10.4 &  2.1 & 17.65 & 23.98 & 21.20 & 19.29 &  0.46 &  0.63 & 17.56 \\
\hline
\end{tabular}
\end{table*}


\subsubsection{Comparison of model-dependent vs.\ model-independent
parameters} \label{s_mdmi}

A check for the accuracy of the model fitting results can be made by
comparing model-independent (MI) with model-dependent (MD) parameters. In
general, there is a very good agreement between our MI and MD parameters,
except for a few particular objects which we discuss below. Isophotal radii
are slightly underestimated by the S\'ersic fits ($\langle
\rho_{27}^\mathrm{S} - \rho_{27} \rangle = -0.6$ arcsec, rms$=1.6$ arcsec),
while the opposite is true for effective radii ($\langle \re^\mathrm{S} -
\re \rangle = 0.7$ arcsec, rms$=0.8$ arcsec). Isophotal magnitudes and
surface brightnesses show very good agreement between their MI and MD
versions, with rms dispersions of 0.11 mag and 0.13 $\msq$,
respectively. Slightly larger differences are obtained for mean effective
surface brightnesses ($\left\langle\langle\mu_\mathrm{e}^\mathrm{S} \rangle
- \sbep\right\rangle = 0.11$, rms$=0.16 \msq$).

The previous results show that, in general, our fits represent a good match
to the observed SBPs, and that the $\mu{(g)} = 27 \msq$ isophote is
appropriate for measuring global photometric parameters. However, systematic
differences are present between MI and MD parameters; they become
particularly evident when comparing isophotal ($g_{27}$) vs.\ integrated
($g_\mathrm{T}$) magnitudes. While the agreement is good for the brightest
galaxies, fainter objects, which have larger fractions of their luminosities
below $\mu{(g)} = 27 \msq$, tend to have $g_{27}-g_\mathrm{T} > 0$. This
effect can be quantified by the outer fraction of galaxy light beyond the
isophotal radius, which can be defined following \citet*{TGC01}:
%
\begin{equation}
\label{e_frfin}
F(\rho^\mathrm{S}_{27}) = {L_\mathrm{T} - L(\rho^\mathrm{S}_{27}) \over
L_\mathrm{T}} = 1 - {\gamma\left[{2/ n}, b_n \left({\rho^\mathrm{S}_{27}
/ \rho^\mathrm{S}_\mathrm{e}}\right)^n\right] \over \Gamma({2/ n})} ,
\end{equation}
%
where $L_\mathrm{T}$ and $L(\rho^\mathrm{S}_{27})$ are the luminosities
integrated from the S\'ersic law, up to infinity and up to the $\mu{(g)} =
27 \msq$ isophote, respectively, while $\Gamma$ and $\gamma$ are the Gamma
and Incomplete-Gamma functions, respectively. Note that $\re^\mathrm{S} \,
b_n^{-1/n} =\rho_\mathrm{o}$, where $b_n$ depends only on $n$.

\begin{figure}
\includegraphics[width=\hsize]{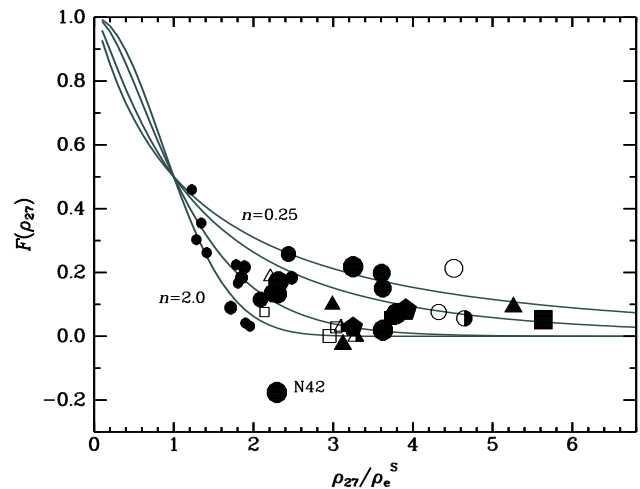}
\caption{Fraction of galaxy light beyond the isophotal radius vs.\ isophotal
to effective radius ratio. Circles: E--dE; pentagons: S0--dS0; squares:
Sa--Sc; triangles: Sd--Im. Filled symbols: definite members; open symbols:
background objects; half-filled symbols: dubious cases. Symbol sizes are
proportional to integrated magnitudes. Curves obtained from
equation~(\ref{e_frfin}) are shown for $n=0.25$, $n=0.4$, $n=1$, and $n=2$.
Note the outlier object (N42 in our catalog) at $F(\rho_{27})=-0.18$, as
discussed in the text.
}
\label{f_frfin}
\end{figure}

In Figure~\ref{f_frfin} we plot $F(\rho_{27}$) against
$\rho_{27}/\re^\mathrm{S}$ for our galaxies (i.e., using MI values for the
isophotal parameters), with filled symbols for definite members, hollow
symbols for background objects, and half-filled symbols for dubious cases,
and coding for different morphological types as indicated in the caption
(see Table~\ref{t_clas}). Also shown are curves obtained from
equation~(\ref{e_frfin}) for different values of $n$ spanning the observed
range. There is a clear agreement between the observations and the expected
$F(\rho_{27})$ values; in particular, the faintest, very LSB objects, have
$\rho_{27}/\rho^\mathrm{S}_\mathrm{e} \la 1.5$ and hence a large
$F(\rho_{27})$.  Isophotal magnitudes thus underestimate the total
luminosities of the faintest galaxies by up to $\sim 0.5$ mag. This should
be taken in mind in subsequent analyses.

In addition, a few bright members show significant differences 
in their MD and MI parameters. This is the case, for instance, of 
galaxy N42, a clear outlier in Fig.~\ref{f_frfin}, with $F(\rho_{27})=-0.18$.
This dS0 galaxy consists in fact of a bulge and disc components and cannot
be easily fitted by a single S\'ersic law. In this case, our integrated 
magnitude only refers to the disc component. We will return on this object in
Sec.~\ref{s_disc}.

\subsection{The surface brightness -- magnitude relation}
\label{s_sbmr}

\begin{figure}
\includegraphics[width=\hsize]{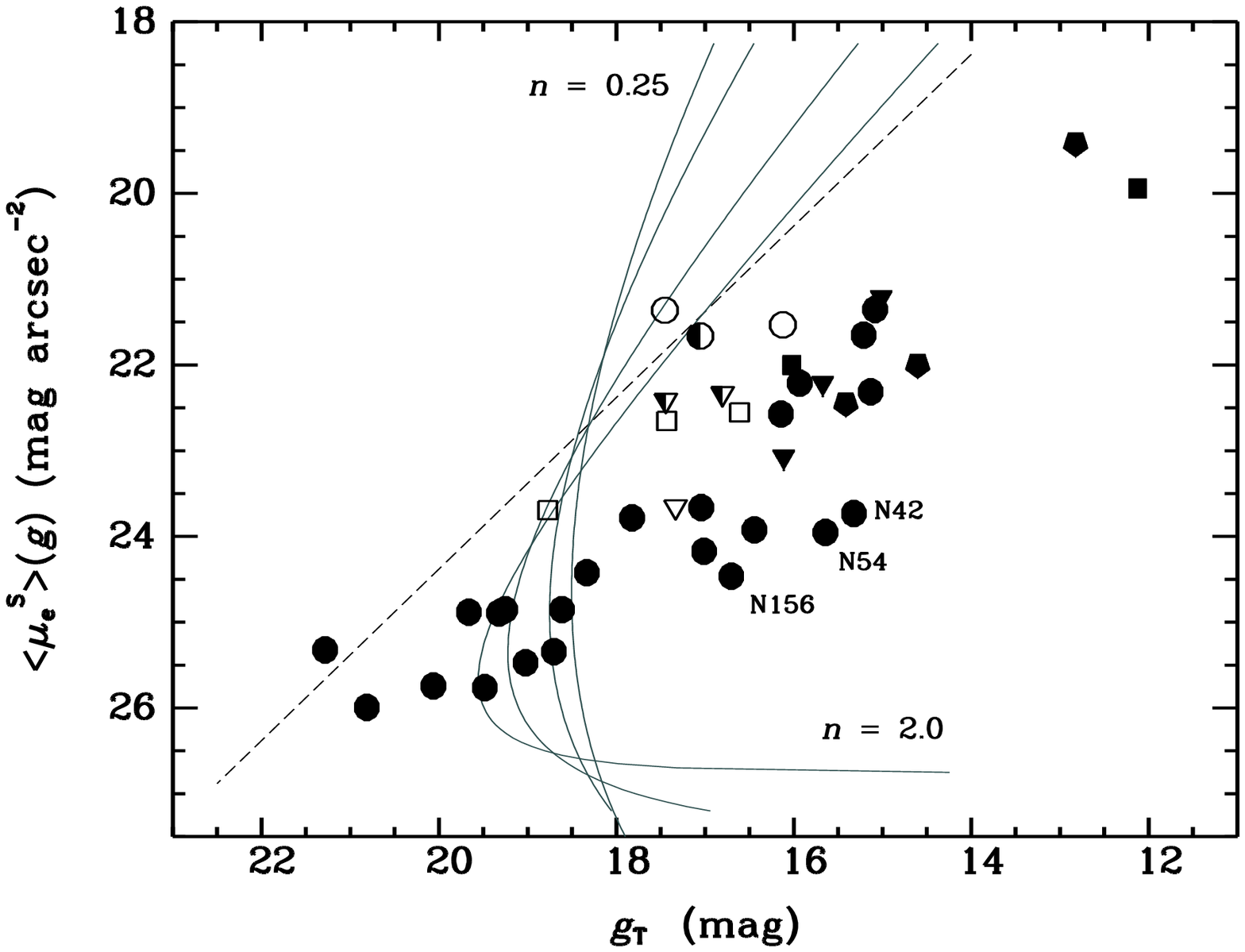}
\caption{Mean effective surface brightness vs.\ integrated magnitude, both
obtained from the model fits in the $g$ Gunn band. Symbol coding is the same
as for Fig.~\ref{f_frfin}. Dashed line: constant effective radius,
$\rho^\mathrm{S}_\mathrm{e} = 3$ arcsec. Solid lines: survey limits in
\citetalias{FS90} (equivalent to $\rho=8$ arcsec at $\mu_{(B)} = 27 \msq$),
for $n=0.25$, $n=0.4$, $n=1$, and $n=2$. Galaxies N42, N54,
  and N156 (see text) are
labelled. Photometry has been corrected for Galactic reddening according to
\citet{BH82}.}
\label{f_sbgt}
\end{figure}

Figure~\ref{f_sbgt} shows the mean effective surface brightness ($\sbep$)
vs.\ integrated magnitude ($g_\mathrm{T}$) diagram for our photometric
sample, with similar symbol coding as in Fig.~\ref{f_frfin}. 
Photometry has been corrected for Galactic reddening according to
\citet{BH82}.\footnote{We adopted $E(B-V) = 0.03$ for the region around
\ngc, which translates into a similar $E(g-r)$ color 
excess, recalling that $A(g)/E(B-V) = 3.55$ and $A(r)/E(B-V) = 2.45$
\citep[after][]{LB82}. For our galaxies, therefore, extinction in the 
$g$ band amounts to $A(g) \sim A(V) \sim 0.11$~mag, in average.} 
The well known
magnitude -- surface brightness relation is evident, with no clear
segregation between early- and late-type galaxies, at least within the narrow
magnitude range where there is overlap.

Confirmed and probable background galaxies occupy a differentiated region in
the diagram, with a higher surface brightness for a given integrated
magnitude. This corresponds to a more ``spiky'' apparent morphology and a
lower effective radius; the $\rho^\mathrm{S}_\mathrm{e} = 3$ arcsec locus
(dashes) is shown as a guideline. However, the distinction between members
and non-members is not clear-cut; any constant radius line, within a
relatively broad range, will leave some members and non-members on each
side. This is particularly true for N109 which, being in the near background
($v_r= 5409 \kms$), lies intermingled with definite members in the
diagram. Hence, only a coarse (statistical) membership classification can be
made through the surface brightness -- magnitude relation
\citep*[e.g.,][]{KDG03}.

A few galaxies scatter below the main trend. Most of these (N54, N70, N83,
N156) are dEs with very shallow SBPs; they have $n \simeq 1.0$ and central
surface brightnesses fainter than 23 $\msq$ in $g$. Galaxy N54 was originally
classified as ``dE(Huge)'', following the designation given by \citet{SB84}
to large-size, LSB dwarfs found in Virgo. The other galaxies might be less
extreme examples of these relatively bright but very LSB dwarfs.  In turn,
N156 shows an additional very interesting feature: a warped disc taking the
form of a pair of LSB outer spiral arms or tidal tails.\footnote{The feature 
is only marginally evident from the contour map of Fig.~\ref{f_conto} but 
it clearly stands out from the inspection of the
original frames at the telescope.} Again, N42 lies off the main $\sbep - g_\mathrm{T}$ 
relation due to its poor S\'ersic fit \citep[see also][]{CB01}.  We
shall be back in more detail to these three notable galaxies 
(i.e.\ N54, N156 and N42) in Sec.~\ref{s_disc}.

Also shown in Fig.~\ref{f_sbgt} as solid lines are the curves of constant
radius ($\rho=8$ arcsec) at the limiting isophote ($\mu(B) = 27 \equiv
\mu(g) \simeq 26.4 \msq$) in the original catalogue of \citetalias{FS90},
for different values of the shape parameter $n$. The points to the left of
the $n=1$ curve correspond to six dSph candidates discovered in the present
work or in \citetalias{C99}\footnote{The seventh dSph, N93A lies just
rightwards the $n=1$ curve, while the SBP of N49A has too low S/N ratio for
any reliable fit to be made.}, plus the very faint N55. Note that selection
effects discriminate against faint objects with low $n$ (i.e., ``concave'')
SBPs, an effect that might be relevant when studying the correlations of $n$
with luminosity or size \citep[see also][]{CFG94}.
 The existence of faint, $n<1$ dwarfs has been confirmed from Local Group
data, where selection effects of a different nature take place
\citep*{JBF00}.

Qualitatively similar graphs are obtained by plotting any combination of
surface brightness vs.\ magnitude, irrespective of whether they be isophotal
or effective, MI or MD. However, the version used here provides the clearest
member--background discrimination, and allows constant limiting radii curves
to be drawn.

\section{Dwarfs with a transition-type morphology}
\label{s_disc}

Since long, observational data have provided ambiguous evidence on the
structure of dwarf elliptical galaxies, whether spheroidal or disky. For
example, while their structureless isophotes made dEs gain their
``elliptical'' designation, their approximately exponential SBPs seemed to
link them to disc systems. Their apparent flattening distribution, in turn,
put dEs in an intermediate situation between normal E and disc galaxies
\citep[e.g.][and references therein]{FB94}.

Recently, kinematical evidence has been presented showing that at least some
dEs in Virgo and the \ngc\ Group are rotationally flattened spheroids
\citep{dRDZH01, PGCSG02}. The non-detection of rotation in a sample of
fainter dEs, led \citet*{GGvdM02} to suggest an association between the
presence of rotation and dE luminosity.  Photometric observations have also
revealed ``hidden'' discs and/or spiral--bar features in a few dEs in Virgo
\citep*{JKB00, BBJ02} and Fornax \citep{dRDZH03}. These results seem to
support models in which present-day dEs in clusters are the remnants of
``harassed'' disc galaxies \citep*{MLK98}.

On the other hand, the evolutionary connection between different dwarf types
(dE -- dI -- BCD) has been a subject for debate since long
\citep[e.g.,][]{SB84, PT96}, hence there is sustained interest in possible
``transition'' objects, since they may hold important keys for dwarf
evolution \citep[e.g.,][]{VSV84, SH91, CB01, SCM03}.
At least three galaxies in our sample, classified as dE or dS0 in
\citetalias{FS90}, show photometric evidence for disc structure. We describe
them in the following subsections.

\subsection{Warped discs in N153 and N156}

N156 was originally classified as dE by \citetalias{FS90}. As discussed in
Sec.~\ref{s_sbmr}, however, the faintest ($\mu(g) \ga 26 \msq$) isophotes in
our images show an ``integral sign'' shape for galaxy morphology, with a
warp or tidal tail emerging at each extreme of the major axis. The $a_4$
Fourier coefficient from the isophote-fitting routine for this galaxy is
slightly negative for $\rho \ga 7$~arcsec, indicating a mild boxiness for
isophotes fainter than $\mu(g) \simeq 25 \msq$.

The low-luminosity spiral N155, a spectroscopically confirmed group member,
lies at a short projected distance (1.5 arcmin 
$\equiv 10.7 \, h_\mathrm{o}^{-1}$
kpc) from the centre of N156. It is thus tempting to invoke an interaction
between both objects as the origin for the warped shape of N156. The facts
that the later's ``arm'' or tidal tail facing N155 points in the right
direction, and that there is a hint for a counter tail at very LSB levels in
N155 itself, seem to support the interaction hypothesis. Note that N155's
mean effective surface brightness is $2.5 \msq$ brighter than that of N156,
implying a higher surface mass density for the former (if similar mass to
light ratios are assumed). Unfortunately, the spectrum we obtained for N156
has an insufficient S/N ratio to derive its radial velocity; hence, its
kinematics with respect to N155 cannot be stated.

Sharing a similar warped appearance, N153 is however a rather different
case. It was originally classified as dS0, because its disky shape is
clearly seen upon visual inspection. In addition, we were able to detect a
warped distortion of the outermost isophotes (see Fig.~\ref{f_conto}). These
features are quantified by the $a_4$ coefficient, which is mildly positive
(i.e., disky) within the range $10 \la \rho \la 20$ arcsec and definitely
negative for $\rho \ga 20$ arcsec, where the warping gives a boxy shape to the
faintest isophotes. This trend is accompanied by a change in ellipticity,
which goes from $\epsilon=0.6$ to  $\epsilon=0.4$ over the same range 
in radius.

N153 is fairly isolated; it has no catalogued companion (irrespective of
membership class) closer than 8.3 arcmin on the sky. One of its nearest
neighbours is, remarkably, the bright spiral NGC\,5054, which lies at $\sim
70\, h_\mathrm{o}^{-1}$ kpc projected distance. However, both objects differ
by more than $1000 \kms$ in radial velocity, hence an interaction is not
likely.

Note that the evidences for disc structure that we found in N153 and N156
are directly observable from their images without any further processing. It
is thus not unreasonable to think that more embedded disc and/or spiral
features may be unveiled after appropriate image processing
\citep[e.g.,][]{BBJ02, dRDZH03}. A search for these kind of structures is
now in progress for a larger sample of early-type dwarfs in the \ngc\ Group
(Cellone \& Buzzoni 2004, in preparation).

\subsection{N42: a bulge plus disc system}

The relevant case of N42 and its anomalous location in Fig.~\ref{f_frfin} 
and \ref{f_sbgt}, has been extensively discussed in previous papers 
(\citetalias{C99}), and especially in \citet{CB01}, where we showed that, 
although a low-$n$ fit produces a nominally better match to the galaxy 
integrated magnitude, its physical sense is questionable. 
Our new data for N42 confirm that, trying a single
S\'ersic law, the resulting fit depends highly on the fitting range in
$\rho$. With the full useful SBP we obtain $n=0.44$, while avoiding the
bulge ($\rho > 15$ arcsec) a convex shape parameter is obtained
($n=1.18$). However, while the lower $n$ value gives a closer match to the
total luminosity of the galaxy (because it tries to include the bulge), both
fits largely overestimate the effective radius of the galaxy as compared to
its MI value.
 In addition, both high- and low-$n$ fits make N42 drop below the main
surface brightness -- magnitude relation (see Sec.~\ref{s_sbmr}).

 In fact, any $n<1$ model leaves, after subtraction from the original image,
both positive and negative (i.e., unphysical) significant residuals.
On the other hand, $n>1$ models give good fits to the outer region of
the galaxy (the disc), leaving, after subtraction, a bulge-type
component about 20\% the total luminosity of the galaxy.
We thus sustain a disc + bulge structure for N42, with the disc being the
more luminous component. We give further support to this conclusion in the
following subsection.

\subsection{The colour -- magnitude diagram}

\begin{figure}
\includegraphics[width=\hsize]{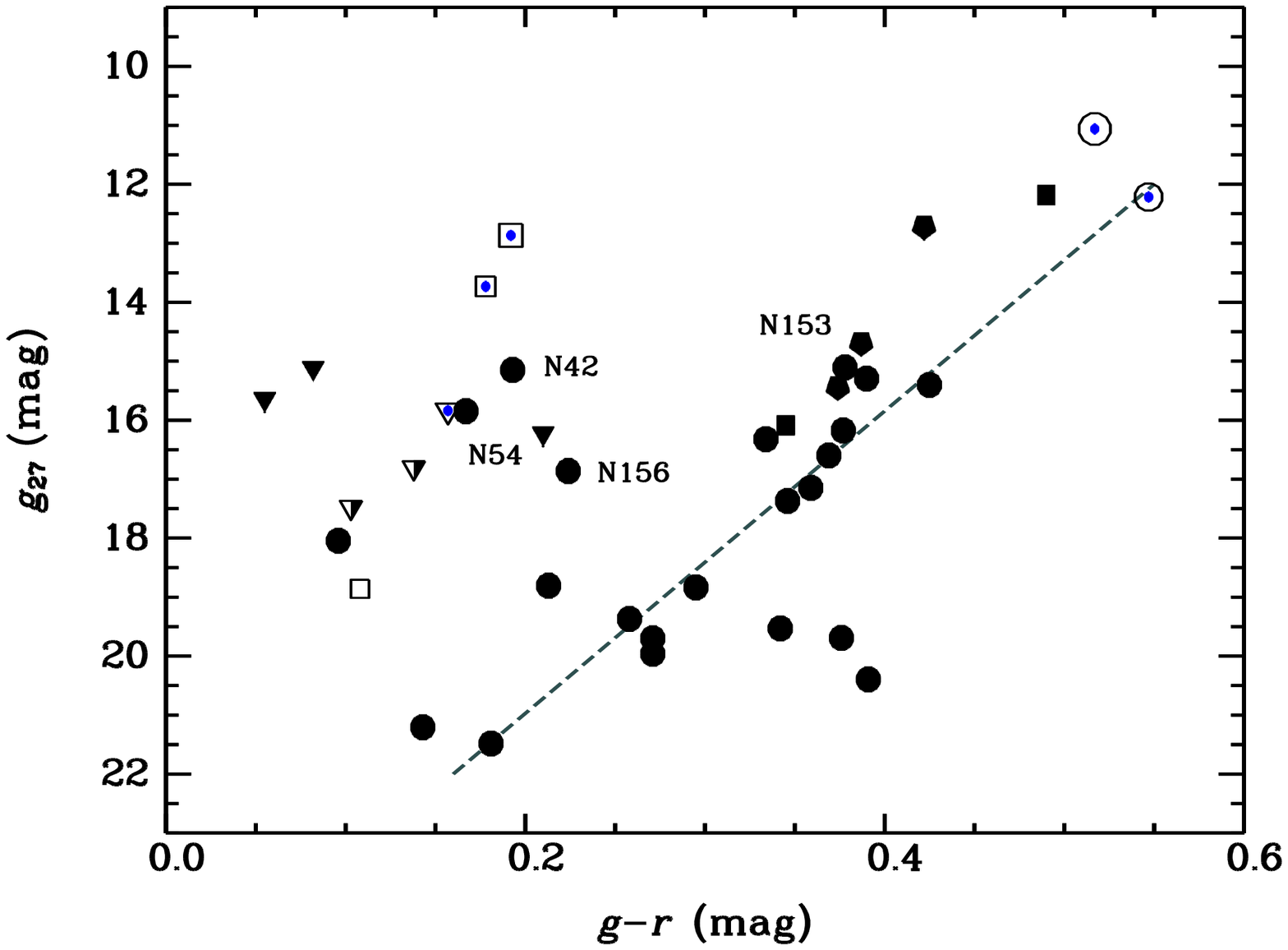}
\caption{Isophotal colour -- magnitude diagram for our sample.  Symbol
coding is the same as for Fig.~\ref{f_frfin}, except that dotted open
symbols correspond to background objects with known redshifts corrected to
the distance of the \ngc\ Group. The dashed line is the $R$ vs.\ $B-R$\,
relation for Coma dEs from \citet{SHP97}, transformed to the Gunn system and
corrected for distance. Data have been corrected for Galactic
reddening according to \citet{BH82}, as explained in Sec.\ 5.2.}
\label{f_colmag}
\end{figure}

Although a thorough study of the evolutionary properties of the
galaxies in the present sample will be presented in a forthcoming
paper (Buzzoni \& Cellone 2004, in preparation), it is interesting to see
now whether the disc-like structural characteristics of the galaxies
discussed in this section show any signature in their stellar
populations.

We thus show in Fig.~\ref{f_colmag} the colour -- magnitude diagram for
all our sample, with the same symbol coding as in Fig.~\ref{f_frfin} and
\ref{f_sbgt}, except that dotted open symbols correspond to
distance-corrected magnitudes for background objects.\footnote{Note
that, for the illustrative scope of Fig.~\ref{f_colmag}, $k$
correction has only been taken into account for background galaxies,
according to \citet{B95}. At the distance of
the \ngc\ Group, $k \simeq 2.5\,\log (1+z) \lesssim 0.01$ mag, that
is $k$ correction is dominated by the geometrical term  
and is therefore negligible in our photometric bands.}
The well known trend
of redder colours (i.e., higher metallicities) for brighter systems
\citep[e.g.,][]{C83,CFG94,SHP97,HKRIQ99} is clearly seen for the early-type
galaxies. Note that, after correcting to the \ngc\ Group distance, the two
confirmed background ellipticals N39 and B3 lie not far from the extrapolation
of the colour -- magnitude relation for dwarfs.

Late-type objects, in turn, form a disjoint sequence with a bluer mean
colour, showing no clear trend with luminosity. Note that the faintest
dwarfs span a broad range in colour, which is probably due in part to large
photometric errors, but may be also due to a spread in mean ages and/or
metallicities. This last point is consistent with the known difficulty to
distinguish between elliptical an irregular dwarfs at the faintest
luminosity levels \citep{FB94}.

Notably, both N42 and N156 lie on the late-type branch, as should be
expected from their disc-dominated structure. The same is true for N54, a
``huge'' dE already mentioned because of its departure form the main trend
in the surface brightness -- magnitude relation (sect.~\ref{s_sbmr}).
Except for the outer warped isophotes in N156, this dwarf and N54 have very
similar structural properties: both are very LSB, $n=1.0$ systems, lying
well within known ranges in size, luminosity and surface brightness of dwarf
irregulars \citep[e.g.,][]{PT96}.  However, there is no hint for zones of
current/recent star formation in these two galaxies (neither in N42);
instead, they have a very smooth appearance, which surely led to their
original dE classification. Hence, we prefer an intermediate early/late type
classification for them. Neutral hydrogen 21 cm observations would be highly
desirable for these objects.

On the other hand, both N70 and N83, which are structurally similar to N54
and N156 (see sect.~\ref{s_sbmr}), lie on the colour -- magnitude relation
for early-type galaxies. Their respective locations in a two-colour diagram
will give some clues on their evolutionary status (Buzzoni \& Cellone 2004, in
preparation)

Finally, N153 also lies on the early-type branch of the colour -- magnitude
relation, despite of its disky structure above discussed. However, its low
shape parameter ($n=0.53$) shows that, despite its detectable disc, N153 is
indeed a bulge-dominated system, with its integrated red colour being the
consequence of a relatively high metallicity.

\section{Summary and Conclusions}

In this paper we presented systematic multicolor photometry for an extended
sample of 33 dwarf and intermediate-luminosity galaxies in the group of
\ngc\ (including observations for the E galaxy \ngc\ itself, and the other
luminous Sb member NGC~5054). For 13 of these objects, also mid-resolution
spectroscopy was collected (this nearly doubles the galaxy sample covered in
the literature for this group) in order to derive full kinematical
information for each individual target and assess the dynamical status for
the \ngc\ Group as a whole.

The Group appears clearly defined in redshift space, with a mean
heliocentric radial velocity, $\langle v_r \rangle = 2461 \pm 84 \kms$ ($z =
0.0082$), and a moderate dispersion, $\sigma_{v_r} = 431 \kms$.  We also
found marginal evidence of a possibly related  sub-structure
(at least four galaxies spectroscopically confirmed) at $v_r \sim 5000-6000
\kms$, while three other galaxy aggregates seem to project on the background
of the \ngc\ field, respectively at $z = 0.045, 0.09$ and 0.28.

Our kinematical data show no luminosity segregation among the early-type
galaxy sub-samples: both the dwarf and bright E/S0 populations show nearly
identical velocity distributions ($\sigma_{vr} \sim 294~\kms$ for dwarfs and
$\sigma_{vr} \sim 287~\kms$ for bright ellipticals), while late-type galaxy
distribution is on the contrary \textbf{sensibly} broader, with $\sigma_{vr}
\sim 680~\kms$.

On the basis of the \griz imagery, and thanks to excellent seeing conditions
of our observations, we tried a revised morphological and membership
classification for the galaxies in the sample. We were able to confirm all
but one (i.e.\ galaxy N109) of the ``definite members'' included in the
spectroscopic subsample, which were originally classified based on
morphological criteria; however, an important fraction of background
galaxies is probably present among ``likely'' and ``possible'' members.

The presence of a nucleus, defined as a pointlike source clearly standing
out from the inwards extrapolation of the galaxy profile, could be detected
in just five out of the nine galaxies originally classified as dE,N, thus
confirming the intrisic difficulty of nuclei identification on photographic
plates.

Deep surface photometry down to $\mu(g) \sim 27 \msq$ allowed us
to detect six new dSph candidates \citepalias[plus two other originally
reported by][and confirmed here]{C99}, most of them at small
projected distances from \ngc, the central galaxy of the Group.

Clear evidence for disc structures in at least three galaxies previously
catalogued as dE or dS0 (namely, N153, N156 and N42), was also obtained.
The fact that this was inferred through different kinds of evidences, not
always simultaneously present in all objects, suggests that these galaxies
probably conform an heterogeneous set. It is thus necessary to extend this
study, in order to shed some light on the evolutionary scenarios that led to
the present-day dE population in groups.  The \ngc\ Group is particularly
suited since it has about half the members as the more studied Fornax
Cluster, however its central density and velocity dispersion are larger. It
would be thus interesting to test whether proposed models to transform discs
into spheroids, like galaxy harassment, could work in different
environments.

In a second paper of this series (Buzzoni \& Cellone 2004, in preparation) we
will further extend our study of the \ngc\ Group trying a more 
specific analysis of the distinctive properties and the evolutionary status of 
stellar populations in LSB galaxies of this Group.

\section*{Acknowledgments}

This work was based on observations collected at the European Southern
Observatory, La Silla (Chile).  It is a pleasure to acknowledge Michael
Sterzik, Martin K\"urster and the whole ESO technical staff for invaluable
support during the observing runs at La Silla.  We wish also to thank
Michael Drinkwater, the referee of this paper, for his constructive
remarks. This project received partial financial support from the Italian
MIUR under COFIN'00 02-016 and CNR/GNA grants for visiting
scientists. Argentinian CONICET is also acknowledged for financial
support. SAC wishes to thank the kind hospitality from the Osservatorio
Astronomico di Brera (Italy), where part of this work was carried out.  This
work made use of the NED database supported at IPAC by NASA.

\end{document}